\documentclass[pre,twocolumn,showpacs]{revtex4}

\usepackage{graphicx}
\usepackage{amsmath}
\usepackage{mathtools}
\usepackage{amssymb}

\newcommand{\ba}{\begin{eqnarray}}
\newcommand{\ea}{\end{eqnarray}}
\newcommand{\be}{\begin{equation}}
\newcommand{\ee}{\end{equation}}

\newcommand{\ra}{\rho({\bf r})}
\newcommand{\rb}{\rho({\bf r}')}
\newcommand{\rc}{\rho({\bf r}'')}
\newcommand{\ua}{u({\bf r},{\bf r}')}

\newcommand{\ha}{h({\bf r},{\bf r}')}

\newcommand{\inta}{\int d{\bf r}}
\newcommand{\intb}{\int d{\bf r}'}
\newcommand{\intc}{\int d{\bf r}''}
\newcommand{\fid}{k_BT\int d{\bf r}\,\rho({\bf r})\big[\log\rho({\bf r})\Lambda^3-1\big]}

\begin{document}

\title{Density functional formulation of the Random Phase Approximation for inhomogeneous fluids: 
application to the Gaussian core and Coulomb particles}

\author{Derek Frydel}
\affiliation{Institute for Advanced Study, Shenzhen University, Shenzhen, Guangdong 518060, China\\
School of Chemistry and Chemical Engineering, 
Shanghai Jiao Tong University, Shanghai 200240, China}
\author{Manman Ma}
\affiliation{Institute of Natural Sciences, Shanghai Jiao Tong University, Shanghai 200240, China}

\date{\today}

\begin{abstract}
Using the adiabatic connection, we formulate the free energy in terms of the correlation function
of a fictitious system, $h_{\lambda}({\bf r},{\bf r}')$, where $\lambda$ determines the interaction strength.
To obtain $h_{\lambda}({\bf r},{\bf r}')$ we use the Ornstein-Zernike equation, and the two equations 
constitute a general liquid-state framework for treating inhomogeneous fluids.  As the two equations do not
form a closed set, an approximate closure relation is required and it determines a type of an approximation.  
In the present work we investigate the random phase approximation (RPA) closure.   
We determine that this approximation is identical to the variational Gaussian approximation
derived within the framework of the field-theory.  
We then apply our generalized RPA approximation to the Gaussian core model and Coulomb charges.  



\end{abstract}

\pacs{
}

\maketitle

\section{Introduction}

Pair interactions of hard-sphere fluids derive from the excluded volume effects:  non-overlapping
configurations recover an ideal-gas behavior, but the exclusion of overlapping configurations reduces 
available phase-space, leading at high density to phase transition.  
In this sense the hard-sphere fluids constitute a geometric problem.  
Within various successful (nonlocal) density functional theories (DFT), 
a free energy functional for hard-sphere fluids is built from a weighted rather than local density --- 
non-locality is attained by construction \cite{Evans79}.  In early prescriptions, a weighted density
corresponded to a convoluted local density, where the single convoluting function was the Mayer 
f-function.  The resulting 
theories, their refinements and extensions came to be known as the weighted DFT theories.  A crucial 
next development was to decompose a Mayer f-function into several weight functions, yielding multiple
weighted densities and, by the same token, muiltiple building blocks from which an approximate $F_{\rm ex}$ 
was to be constructed \cite{Rosenfeld89,Tarazona08,Evans09,Roth10,Frydel12}.  Referred to as the fundamental 
measure (FM) DFT, a nice feature of this approach is the capture of a correct dimensional crossover:  
each consecutive 
reduction of the system dimensionality, 3D$\to$2D$\to$1D$\to$0D, recovers either an accurate  
or exact $F_{\rm ex}$.  

The success of the hard-sphere DFT theories (and the lack of equivalent theories for arbitrary pair 
interactions), prompted attempts to implement the hard-sphere framework to 
other types of short-range interactions.  It became something of a standard method to map particles 
with arbitrary short-range interactions onto a hard-sphere fluid by ascribing to a pair potential of interest  
an effective diameter.  Density profiles are then obtained from 
the DFT theories for hard-spheres.  The Barker-Henderson effective diameter is one recipe among others 
for extracting an effective diameter \cite{Barker67}.   

A more sophisticated example is the "soft" fundamental 
measure DFT developed for penetrable spheres (spheres may overlap but at an energy cost). 
Within this method $F_{\rm ex}$ is constructed to satisfy a correct 
dimensional crossover \cite{Schmidt99a,Schmidt99b,Schmidt00,Lowen02,Sweatman02}.  

But for particles with arbitrary pair interactions, where excluded volume effects are not dominant, 
the mean-field approximation is still a preferred theoretical tool 
\cite{Hansen00,David07,Frydel11,Frydel13,Frydel15a}.  An obvious example are charged particles with
long-range interactions.  Other examples are particles with bound (non-divergent) interactions, known as 
penetrable particles.  This class includes the Gaussian core model, or the already mentioned penetrable 
spheres.  
But the correlations neglected by the mean-field description are not always trivial. This is particularly true
of Coulomb systems.  In such a case the "beyond-mean-field" approach is desirable.  
Splitting the excess free energy into the mean-field and correlation contribution, 
$F_{\rm ex}[\rho]=\frac{1}{2}\int d{\bf r}\int d{\bf r}'\,\ra\rb\ua + F_c[\rho]$, where $\ra$ is a number density, 
and $\ua$ is an arbitrary pair interaction, the "beyond-mean-field" approach amounts to finding an appropriate  
functional $F_c[\rho]$.

The "beyond-mean-field" approximations for Coulomb systems are dominantly formulated within the field-theoretical 
framework based on mathematical transformation of a partition function, using a Gaussian integral identity 
\cite{Frydel15}, into a functional integral over an 
auxiliary fluctuating field \cite{Rudi88,Attard88,Duncan92,Netz00,Netz03,Wang10,Hatlo14,Tony14,Frydel15}.  
The saddle-point of the effective Hamiltonian recovers the mean-field solution, while the harmonic 
fluctuations around the saddle-point account for weak (Gaussian) correlations.  If formulated variationally, 
the equations become self-consistent (non-perturbative) and generally are deemed superior
to the perturbative formulation \cite{Netz03,Wang10,Frydel15}.  

A drawback of the field-theoretical formulation is the loss of physical intuition after one 
moves from physical to auxiliary phase-space.  In the present work we re-derive the variational Gaussian 
equations of the field-theoretical framework using only the liquid-state theory.  The Gaussian 
approximation is equivalent to the well established random phase approximation (RPA) 
\cite{Pines52}, a mathematical signature of which is its being comprised of an infinite 
summation of ring diagrams \cite{Mann57}.  
Our formulation of the RPA is general 
and in principle applicable to any pair interactions.  We apply our generalized RPA approximation
to the Gaussian core model, considered to be a weakly correlated fluid \cite{Hansen00}, 
a one-component plasma, and finally a symmetric electrolyte.  

In Sec. \ref{sec:free-energy} we formulate the free energy within the liquid-state formalism 
using the adiabatic connection. By coupling it to the OZ equation, we set up a general theoretical 
framework for inhomogeneous fluids.  
In Sec. \ref{sec:RPA} we consider the RPA closure and derive the appropriate self-consistent 
equations.  
In Sec. \ref{sec:multiple} we generalize the equations to multiple species. 
Finally, in Secs. \ref{sec:GCM}, \ref{sec:OCP}, and \ref{sec:electrolyte} we test the RPA approximation
on concrete systems with wall geometry.

\section{Adiabatic connection formulation of the free energy}
\label{sec:free-energy}
Given a general Hamiltonian for a system of interacting particles, 
\be
H_{} = \sum_{i=1}^N U_{}({\bf r}_i) + \frac{1}{2}\sum_{i\ne j}^N u({\bf r}_i,{\bf r}_j),
\label{eq:H_aux}
\ee
where $U({\bf r})$ is an external potential, $\ua$ is a pair interaction, and $N$ is the number of 
particles, our aim is to obtain a free energy expression in terms of physically meaningul quantities.  
To this end we use the adiabatic connection route \cite{Perdew77,Ren12}, wherein 
interactions are gradually switched on within a fictitious $\lambda$-dependent system, 
\be
H_{\lambda} = \sum_{i=1}^N U_{\lambda}({\bf r}_i)
+ \frac{\lambda}{2}\sum_{i\ne j}^N u({\bf r}_i,{\bf r}_j), 
\label{eq:H_aux}
\ee
where the $\lambda$-dependent external potential $U_{\lambda}({\bf r})$ is introduced to keep a density 
fixed at its physical value for all $\lambda$, and $\lambda=1$ recovers the physical potential, 
$U_{\lambda=1}({\bf r}) = U({\bf r})$.  

The partition function and the free energy of a fictitious system are
\be
Z_{\lambda} = \int d{\bf r}_1\dots\int d{\bf r}_N\,e^{-\beta H_{\lambda}}
\ee 
and
\be
\beta F_{\lambda} = -\log Z_{\lambda},
\ee
respectively.  The free energy of a physical system can be expressed in terms of a fictitious system, 
as a thermodynamic integration, 
\be
F = F_0 + \int_0^{1} d\lambda\,\frac{\partial F_{\lambda}}{\partial\lambda},
\label{eq:FTI}
\ee
where the reference free energy is 
\be
F_0[\rho] = F_{\rm id}[\rho] + \inta\,\ra U_{\lambda=0}({\bf r}),
\ee
and 
\be
F_{\rm id}[\rho] = \fid
\ee
is an ideal-gas free energy.  
The integrand in Eq. (\ref{eq:FTI}) can be written as
\ba
\frac{\partial F_{\lambda}}{\partial\lambda} 
&=& \inta\,\ra\frac{\partial U_{\lambda}({\bf r})}{\partial\lambda} \nonumber\\
&+& \frac{1}{2}\inta\intb\,\ra\rb\ua\nonumber\\
&+& \frac{1}{2}\inta\intb\,\ra\rb h_{\lambda}({\bf r},{\bf r}')\ua \nonumber\\
\ea
and Eq. (\ref{eq:FTI}) becomes
\ba
F[\rho] &=& F_{\rm id}[\rho] + \inta\,\ra U_{\lambda=0}({\bf r})\nonumber\\ 
&+& \int_0^1 d\lambda\inta\,\ra\frac{\partial U_{\lambda}({\bf r})}{\partial\lambda}\nonumber\\
&+& \frac{1}{2}\inta\intb\,\ra\rb\ua\nonumber\\
&+&\frac{1}{2}\inta\intb\,\ra\rb\ua \int_0^1 d\lambda\,h_{\lambda}({\bf r},{\bf r}').\nonumber\\
\label{eq:F1}
\ea
Then after a few cancelations we arrive at the final form, 
\ba
F[\rho] &=& F_{\rm id}[\rho] + \inta\,\ra U({\bf r})\nonumber\\ 
&+& \frac{1}{2}\inta\intb\,\ra\rb\ua\nonumber\\
&+&\frac{1}{2}\inta\intb\,\ra\rb\ua \int_0^1 d\lambda\,h_{\lambda}({\bf r},{\bf r}').\nonumber\\
\label{eq:F1}
\ea
Note that the final expression does not depend on the fictitious potential $U_{\lambda}$.  
The only quantity that depends on $\lambda$ is a correlation function $h_{\lambda}({\bf r},{\bf r}')$.
The last line of the expression represents the correlation free energy,
\be
F_c[\rho] = \frac{1}{2}\int_0^1 d\lambda\inta\,\ra\bigg[\intb\,\rb h_{\lambda}({\bf r},{\bf r}')\ua\bigg].
\label{eq:Fc_1}
\ee
Not surprisingly, $F_c$ depends on the correlation function, $h_{\lambda}({\bf r},{\bf r}')$, that is 
obtained from the Ornstein-Zernike equation (OZ),
\be
h_{\lambda}({\bf r},{\bf r}') = 
c_{\lambda}({\bf r},{\bf r}') + \intc\,\rc h_{\lambda}({\bf r}',{\bf r}'')c_{\lambda}({\bf r},{\bf r}''),
\label{eq:OZ}
\ee
which is a well known exact relation within the liquid-state theory.  Because the direct correlation 
function $c_{\lambda}({\bf r},{\bf r}')$ is not known, an appropriate closure relation is still required.  

\section{Random phase approximation}
\label{sec:RPA}
We consider the simplest closure available, 
\be
c_{\lambda}^{\rm }({\bf r},{\bf r}') = -\beta\lambda\ua,
\label{eq:c_RPA}
\ee
known as the random phase approximation (RPA).  
The closure modifies the exact Ornstein-Zernike relation in Eq. (\ref{eq:OZ}),
\ba
h_{\lambda}^{\rm }({\bf r},{\bf r}') &=& 
-\beta \lambda\ua\nonumber\\ 
&-& \beta\lambda\int \!d{\bf r}''\rc h_{\lambda}^{\rm }({\bf r}',{\bf r}'')u({\bf r},{\bf r}'').\nonumber\\
\label{eq:OZ_RPA2}
\ea
Accordingly, we refer to it as the OZ-RPA equation.  
Application of the OZ-RPA modifies the correlation free energy in Eq. (\ref{eq:Fc_1}), 
\be
F_c^{\rm }[\rho] = 
- \frac{1}{2}\inta\ra\int_0^1 d\lambda\,\frac{h_{\lambda}^{\rm }({\bf r},{\bf r})}{\lambda\beta}
-\frac{u(0)}{2}\inta\,\ra.
\label{eq:Fc_2}
\ee
In the above equation $u(0)=u({\bf r},{\bf r})$.  Likewise, for a homogeneous fluids in a bulk we 
write $h_b(0)= h_b({\bf r},{\bf r})$.  

\subsection{Connection with the field-theoretical formulation}
The $\lambda$-dependence in $F_c$ can be eliminated by expanding
$h_{\lambda}^{\rm }({\bf r},{\bf r}')$, 
\ba
&&h_{\lambda}^{\rm }({\bf r},{\bf r}') =-\beta\lambda u({\bf r},{\bf r}') \nonumber\\
&+&\beta^2\lambda^2\int \!\!d{\bf r}_1\,\rho({\bf r}_1) u({\bf r},{\bf r}_1)u({\bf r}_1,{\bf r}')\nonumber\\
&-& \beta^3\lambda^3\int \!\!d{\bf r}_1\!\!\int \!\!d{\bf r}_2\,
\rho({\bf r}_1)\rho({\bf r}_2)u({\bf r},{\bf r}_1)u({\bf r}_1,{\bf r}_2)u({\bf r}_2,{\bf r}')\nonumber\\ 
&+& \dots
\ea
The expansion is generated iteratively 
by repeated insertion of the right hand side of Eq. (\ref{eq:OZ_RPA2}) for every 
occurrence of $h_{\lambda}^{\rm }({\bf r},{\bf r}')$.  The notation is simplified by introducing an operator
\be
A({\bf r},{\bf r}') = \beta\ra\ua,
\ee
and adopting a convention
\be
A^n \!= \!\!\int \!\!d{\bf r}_1\!\!\int\!\! d{\bf r}_2\!\dots\!\!\int\! d{\bf r}_{n-1}A({\bf r},{\bf r}_1)A({\bf r}_1,{\bf r}_2)
\dots A({\bf r}_{n-1},{\bf r}')
\ee
by means of which we get
\ba
\ra h_{\lambda}^{\rm }({\bf r},{\bf r}') &=& -\lambda A + \lambda^2A^2 - \lambda^3 A^3+\dots
\nonumber\\ &=& -\bigg(\frac{\lambda A}{I+\lambda A}\bigg),
\ea
where $I=\delta({\bf r},{\bf r}')$ is the identity matrix in the continuum limit.  
Integration over $\lambda$ now is done explicitly,
\be
\int_0^{1}d\lambda\, \frac{\ra h_{\lambda}^{\rm }({\bf r},{\bf r}')}{\lambda} = 
-A + \frac{A^2}{2} - \frac{A^3}{3} +\dots = -\log[I+A], 
\label{eq:Fc_Tr}
\ee
and $F_c^{\rm }[\rho]$ becomes
\be
F_c^{\rm }[\rho] = \frac{k_BT}{2}{\rm Tr}\,\log[I+A] - \frac{u(0)}{2}\inta\,\ra,
\label{eq:Fc_3}
\ee
where the first term yields an infinite series of ring diagrams, a characteristic feature of the RPA.  

The expression can further be rearranged by using the formal matrix identity, 
\be
\frac{1}{2}{\rm Tr}\,\log[I+A] = \log\sqrt{\det\big[I+A\big]},
\ee
and the fact that a functional determinant is a solution of a Gaussian functional integral,
\be
\frac{1}{\sqrt{\det\big[I+A\big]}} 
= \int {\mathcal D}\phi\,e^{-\frac{1}{2}\inta\intb\,\phi({\bf r})\phi({\bf r}')[\delta({\bf r}-{\bf r}')+A({\bf r},{\bf r}')]},
\ee
where $\phi({\bf r})$ is a fluctuating field and $\int{\mathcal D}\phi$ is a functional integral. 
The partition function within the RPA can now be written as a Gaussian functional integral,
\ba
Z_{\rm rpa} 
&=&e^{\frac{\beta N}{2}u(0)}e^{-\beta F_{\rm mf}}
\nonumber\\&\times&
\int {\mathcal D}\phi\,e^{-\frac{1}{2}\inta\intb\,\phi({\bf r})\phi({\bf r}')[\delta({\bf r}-{\bf r}')+A({\bf r},{\bf r}')]},
\nonumber\\
\ea
where we used $F = F_{\rm mf} + F_{c}$ and $Z = e^{-\beta F_{\rm mf}}e^{-\beta F_c}$.  
The functional integral formulation has been recovered without resorting to the 
Hubbard-Stratonovich transformation, starting from the liquid-state formulation.

\subsection{density profile}
To obtain an equilibrium density we use the known thermodynamic condition,
\be
\frac{\delta F}{\delta\ra} = \mu,
\ee
where $\mu$ denotes the chemical potential.  
The functional derivative of $F_c$ with respect to $\ra$ incidentally eliminates all
$\lambda$-dependence, 
\ba
\frac{\delta F_c}{\delta\ra} &=& 
\frac{k_BT}{2}\frac{\delta{\rm Tr}\,\log\big[I+A\big]}{\delta\ra}-\frac{1}{2}u(0) \nonumber\\
&=&-\frac{1}{2}\Big[u(0) + k_BTh^{\rm }({\bf r},{\bf r})\Big],
\ea  
and the functional derivative is written in terms of a correlation function of a physical system, 
$h({\bf r},{\bf r})$.  The number density that results is
\be
\ra = \rho_b e^{-\beta U({\bf r})}e^{-\beta\intb\,\rb\ua}
e^{\frac{1}{2}[\beta u(0)+h^{\rm }({\bf r},{\bf r})]+\beta\mu_{\rm ex}^{\rm }},
\label{eq:rho_1a}
\ee
where we separated a chemical potential into ideal and excess parts, 
$\mu=\mu_{\rm id}+\mu_{\rm ex}$, with the ideal contribution related to a bulk density,
\be
\rho_b = \bigg(\frac{e^{\beta\mu_{\rm id}}}{\Lambda^3}\bigg).  
\ee
The excess chemical potential within the present approximation is
\be
\mu_{\rm ex}^{\rm } = \rho_b\int d{\bf r}\,u(r) - \frac{1}{2}\big[u(0) + k_BTh^{\rm }_b(0)\big],
\label{eq:mu_ex}
\ee
where $h^{\rm }_b(r)$ is a correlation function in a bulk.  For $U({\bf r})=0$, we accurately recover 
a bulk density, $\ra\to\rho_b$.   More conveniently, a density can be written as
\be
\ra = \rho_b e^{-\beta U({\bf r})}e^{-\beta\intb\,(\rb-\rho_b)\ua}
e^{\frac{1}{2}\big[h^{\rm }({\bf r},{\bf r})-h^{\rm }_b(0)\big]}.
\label{eq:rho_1}
\ee

\subsection{pressure}
Another quantity of interest is pressure that can be obtained from a type of thermodynamic
integration involving a chemical potential,
\be
P_{\rm ex} = \int_0^{\rho_b} d\rho\,\rho\frac{\partial\mu_{\rm ex}}{\partial\rho},
\ee
where $P = k_BT\rho_b + P_{\rm ex}$.   
The resulting expression shows $\lambda$-dependence, 
\be
P_{\rm ex} = \frac{1}{2}\rho_b^2\int d{\bf r}\,u(r) 
- \frac{k_BT}{2}\rho_b \bigg[h^{\rm }_b(0)
-\int_0^{1} d{\lambda}\,\frac{h_{b,\lambda}^{\rm }(0)}{\lambda}\bigg].
\ee

\section{Multiple species}
\label{sec:multiple}
We next generalize the RPA to multiple species.  The fictitious Hamiltonian, 
equivalent to that in Eq. (\ref{eq:H_aux}), is
\be
H_{\lambda} = \sum_{i=1}^N U_i^{\lambda}({\bf r}_i)
+ \frac{\lambda}{2}\sum_{i\ne j}^N u_{ij}({\bf r}_i,{\bf r}_j).
\label{eq:H_aux2}
\ee
Here we assume that each particle feels different external potential, and pair interactions between different
pairs are different.  Of course, particles are not all different but are grouped 
into species.  

Assuming $K$ different species, the free energy from adiabatic connection is
\ba
F[\{\rho_k\}] &=& F_{\rm id}[\{\rho_k\}] + \sum_{k=1}^K\!\int\! d{\bf r}\,U_k({\bf r})\rho_k({\bf r})\nonumber\\ 
&\!\!\!\!\!\!\!\!\!\!\!\!\!\!\!\!\!+& \!\!\!\!\!\!\!\!\!\!\!
\frac{1}{2}\sum_{k,l}^K\!\int\! d{\bf r}\!\int\! d{\bf r}'\,\rho_k({\bf r})\rho_l({\bf r}') u_{kl}({\bf r},{\bf r}')\nonumber\\
&\!\!\!\!\!\!\!\!\!\!\!\!\!\!\!\!\!+& \!\!\!\!\!\!\!\!\!\!\!
\frac{1}{2}\sum_{k,l}^K\!\int\! d{\bf r}\!\int\! d{\bf r}'\,\rho_k({\bf r})\rho_l({\bf r}') u_{kl}({\bf r},{\bf r}') 
\!\int_0^1 \!\!d\lambda\,h_{kl}^{\lambda}({\bf r},{\bf r}'),\nonumber\\
\label{eq:F2}
\ea
where the ideal-gas contribution is
\be
\beta F_{\rm id}[\{\rho_k\}] = \sum_{i=1}^K\inta\,\rho_k({\bf r})\Big[\log\rho_k({\bf r})\Lambda^3-1\Big].  
\ee
If the Ornstein-Zernike equation for multiple-species is
\be
h_{kl}^{\lambda}({\bf r},{\bf r}') = c_{kl}^{\lambda}({\bf r},{\bf r}') + 
\sum_{n=1}^K\!\int \!d{\bf r}''\rho_n({\bf r}'')h_{nl}^{\lambda}({\bf r}',{\bf r}'')c_{kn}^{\lambda}({\bf r}'',{\bf r}),
\ee
where correlations between particles of a species $k$ and $l$ are mediated by all particles disregarding
their type, then the RPA closure, $c_{kl}^{\lambda}=\beta\lambda u_{kl}({\bf r},{\bf r}')$, yields
\ba
h_{kl}^{\lambda}({\bf r},{\bf r}') &=& -\beta\lambda u_{kl}({\bf r},{\bf r}')\nonumber\\ 
&-&\beta\lambda\sum_{n}\intc\rho_n({\bf r}'')
h_{nl}^{\lambda}({\bf r}',{\bf r}'')u_{kn}({\bf r}'',{\bf r}),\nonumber\\
\ea
and the RPA correlation free energy is
\ba
F_c^{\rm rpa} &=& -\frac{1}{2}\sum_{k=1}^Ku_{kk}(0)\inta\,\rho_k({\bf r}) \nonumber\\
&-&\frac{1}{2}\sum_{k=1}^K\inta\,\rho_k({\bf r}) 
\int_0^1 d\lambda\,\frac{h_{kk}^{\lambda}({\bf r},{\bf r})}{\beta\lambda}.
\ea
(Compare with Eq. (\ref{eq:Fc_2}) for a one component system).  The lack of dependence on inter-species 
correlations, that is, $h_{kl}({\bf r},{\bf r}')$ for $k\ne l$, at first glance appears inaccurate.  But as correlations 
between particles of the same species, $h_{kk}({\bf r},{\bf r}')$, are mediated by all the particles disregarding 
their type, the cross-correlations are always implicit in $h_{kk}({\bf r},{\bf r}')$.

\subsection{density}
An equilibrium density of a species $k$ is obtained from the condition
\be
\frac{\delta F}{\delta\rho_{k}({\bf r})} = \mu_k,
\ee
and the correlational counterpart yields
\be
\frac{\delta F_c}{\delta\rho_k({\bf r})} = -\frac{u_{kk}(0)}{2} - \frac{k_BT}{2}h_{kk}^{\rm }({\bf r},{\bf r}). 
\ee
If the excess chemical potential of a specie $k$ is
\be
\mu_k^{\rm ex} = \sum_{l}^K\rho_l^b\int d{\bf r}\,u_{kl}(r) - \frac{1}{2}\Big[u_{kk}(0) 
+ k_BTh_{kk}^b(0)\Big], 
\ee
then a density is
\ba
\rho_k({\bf r}) &=& \rho_k^b e^{-\beta U_{k}({\bf r})} 
e^{-\beta\sum_{l=1}^K\int d{\bf r}'\,\big(\rho_l({\bf r}')-\rho_l^b\big)u_{kl}({\bf r},{\bf r}')}\nonumber\\
&\times&e^{\frac{1}{2}\big[h_{kk}^{\rm }({\bf r},{\bf r})-h_{kk}^b(0)\big]}.
\ea

\subsection{pressure}
To obtain the pressure we use 
$P_{\rm ex} = \sum_{k=1}^K\rho_k^b\mu_k^{\rm ex} - f_{\rm ex}$,
where the excess free energy density in a bulk is
\ba
f_{\rm ex} &=& \frac{1}{2}\sum_{k,l}^{K}\rho^b_k\rho^b_l\inta\,u_{kl}(r) \nonumber\\
&-& \frac{1}{2}\sum_{k=1}^K\rho^b_k \bigg[u_{kk}(0) + k_BT
\int_0^1 d\lambda\,\frac{h^{b,\lambda}_{kk}(0)}{\lambda}\bigg].\nonumber\\
\ea
The excess pressure then becomes
\ba
P_{\rm ex} &=& \frac{1}{2}\sum_{k,l}^K\rho_k^b\rho_l^b\int d{\bf r}\,u_{kl}(r) \nonumber\\
&-& \frac{k_BT}{2}\sum_{k=1}^K\rho_k^b \bigg[h_{kk}^{\rm }(0)
-\int_0^{1} d{\lambda}\,\frac{h_{kk}^{b,\lambda}(0)}{\lambda}\bigg].\nonumber\\
\ea

\section{The Gaussian core model (GCM)}
\label{sec:GCM}
We apply the developed RPA approximation to the Gaussian core model (GCM), 
whose pair interactions have the Gaussian functional form,
\be
\beta u(r) = \varepsilon e^{-r^2/\sigma^2}.
\ee
$\sigma$ is the length scale that determines the interaction range and $\varepsilon$ determines 
the interaction strength.  Because the potential is bound, the GCM particles are said to be 
penetrable.  

For a homogenous system the free energy in Eq. (\ref{eq:Fc_3}) can be calculated exactly, and
each individual ring term becomes
\be
{\rm Tr}\,A^n = 
\frac{V(\varepsilon\eta_b)^n}{(n\pi\sigma^2)^{3/2}}
\ee
where $\eta_b=\pi^{3/2}\sigma^3\rho_b$ is the reduced density and $V$ is the volume of a system.  
The correlation free energy density, $f_c=F_c/V$, becomes 
\ba
f_c
&=& \frac{\varepsilon\rho_b}{2}\sum_{n=2}^{\infty}\frac{(-\varepsilon\eta)^{n-1}}{n^{5/2}}\nonumber\\
&=&-\frac{\varepsilon\rho_b}{2}\bigg\{1+\frac{{\rm Li}_{5/2}[-\varepsilon\eta]}{\varepsilon\eta}\bigg\},
\ea
where 
${\rm Li}_m(x) = \sum_{n=1}^{\infty}\frac{x^n}{n^{m}}$ is a polylogarithm.
We may now obtain any quantity of interest.  For example, the excess chemical potential is
\be
\beta\mu_{\rm ex}^{\rm } = \varepsilon\eta 
- \frac{\varepsilon}{2}\bigg\{1+\frac{{\rm Li}_{3/2}\big[-\varepsilon\eta\big]}{\varepsilon\eta}\bigg\}
\ee
Comparing with Eq. (\ref{eq:mu_ex}) we get another useful quantity,
\ba
h_b^{\rm }(0) 
&=& \varepsilon\sum_{n=1}^{\infty}\frac{(-1)^n(\varepsilon\eta)^{n-1}}{n^{3/2}}, 
\ea
and the pressure is written as
\be
\frac{\beta P_{\rm ex}}{\rho_b} = \frac{\varepsilon}{2}\bigg\{\eta  
+\frac{{\rm Li}_{5/2}(-\varepsilon\eta)-{\rm Li}_{3/2}(-\varepsilon\eta)}{\varepsilon\eta}\bigg\}.
\ee

Our primary interest, however, lies in the RPA as a theory of inhomogeneous fluids.  
Considering a fluid confined by a hard wall at $x=0$ to a half space $x>0$, we can use  
the contact value theorem to predict the density at a contact with a wall from a bulk pressure, 
\ba
\rho(0) &=& \beta P\nonumber\\ 
&=& \rho_b\bigg[1 + \frac{\varepsilon\eta}{2} + 
\frac{{\rm Li}_{5/2}(-\varepsilon\eta)-{\rm Li}_{3/2}(-\varepsilon\eta)}{2\eta}\bigg].\nonumber\\
\ea
The first term is the ideal-gas contribution, the second is the mean-field contribution, 
and the last term accounts for the RPA correlations.  

\begin{figure}[h] 
 \begin{center}
 \begin{tabular}{rr}
  \includegraphics[height=0.2\textwidth,width=0.35\textwidth]{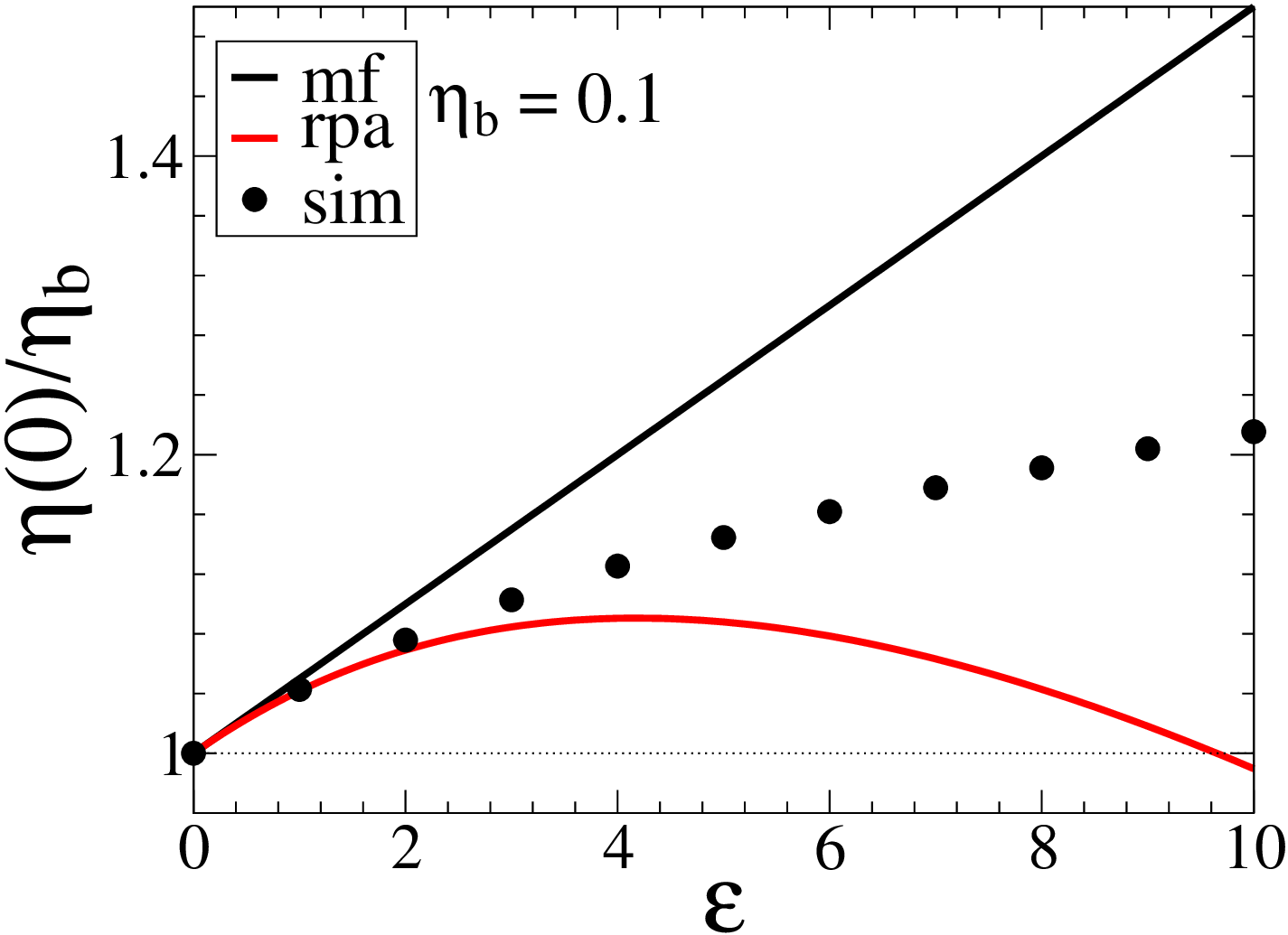}\\
  \includegraphics[height=0.2\textwidth,width=0.35\textwidth]{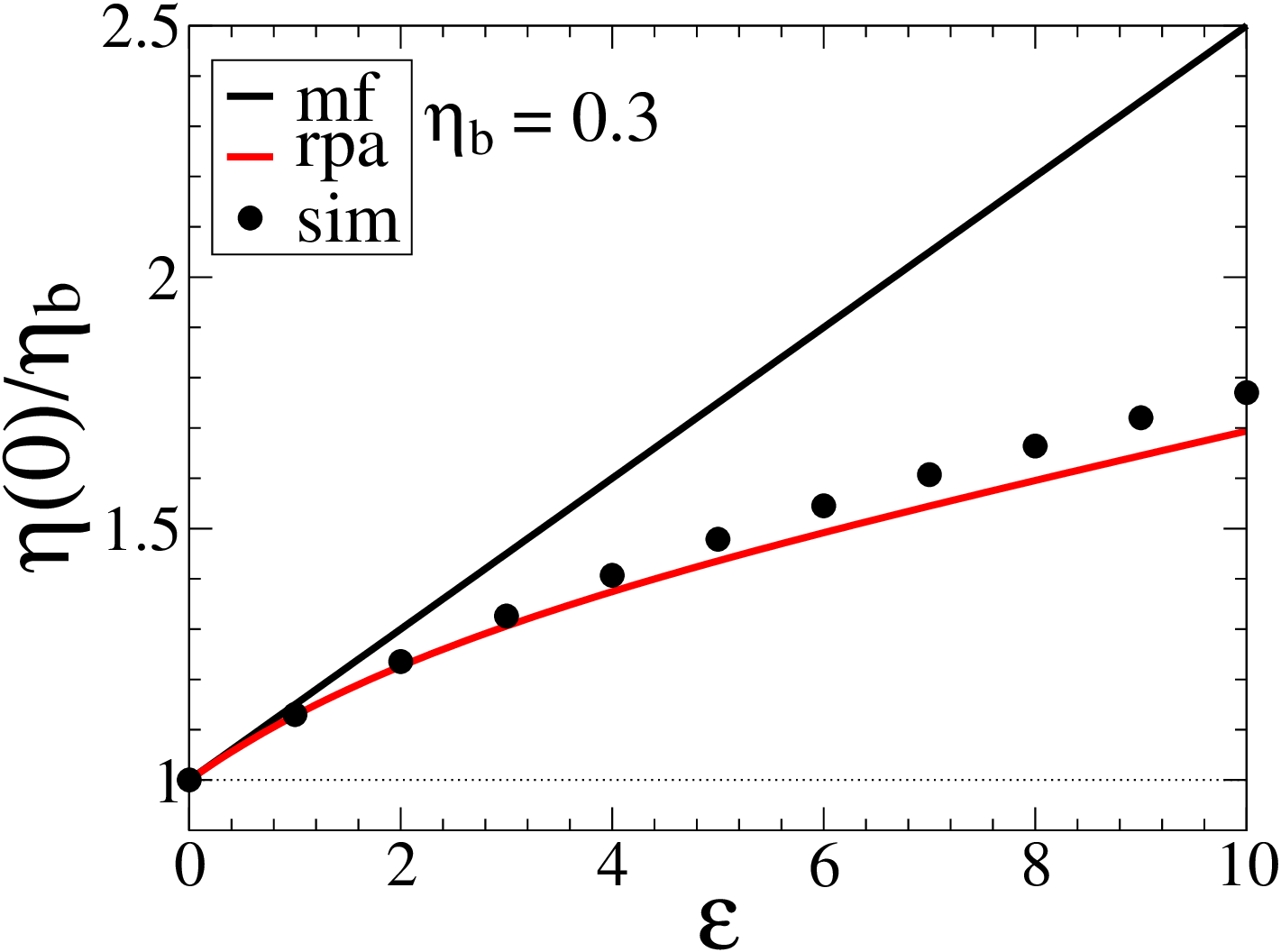}\\
  \includegraphics[height=0.2\textwidth,width=0.35\textwidth]{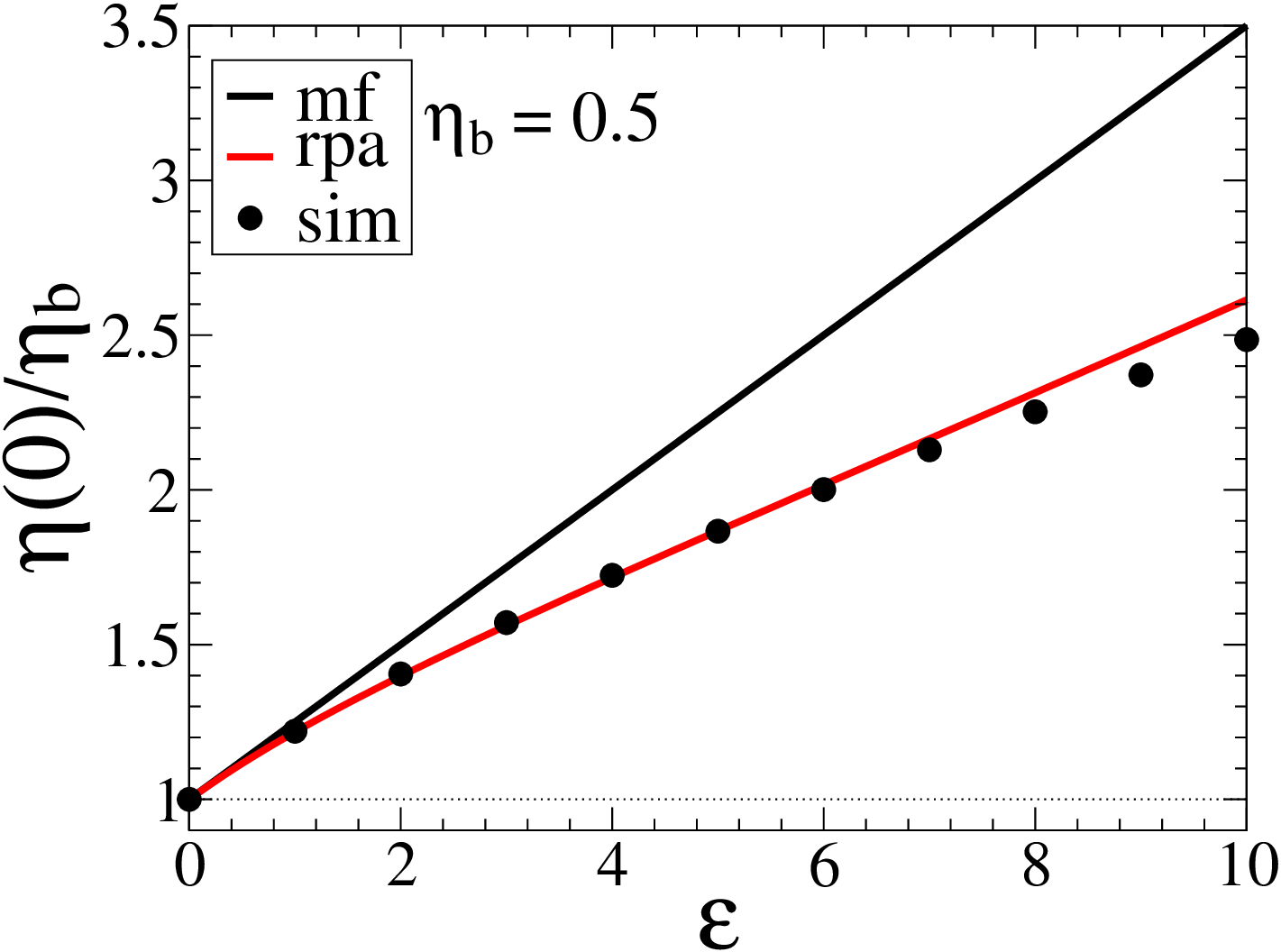}\\
 \end{tabular}
 \end{center}
\caption{The contact density, $\eta(0)=\pi^{3/2}\sigma^3\rho(0)$, as a function of an interaction strength 
$\varepsilon$ for a one component GCM.  $\eta_b$ is the bulk reduced density.  
The dotted horizontal line corresponds to an ideal-gas
prediction.  }
\label{fig:eta_1}
\end{figure}

In Fig. (\ref{fig:eta_1}) we compare the contact density at a wall as a function of $\varepsilon$ for different 
approximations.  The RPA correlations lower the mean-field predictions, and the RPA corrections
become more accurate at high densities, that is, for a larger number of overlaps.  
\graphicspath{{figures/}}

\begin{figure}[h] 
 \begin{center}
 \begin{tabular}{rr}
  \includegraphics[height=0.35\textwidth,width=0.45\textwidth]{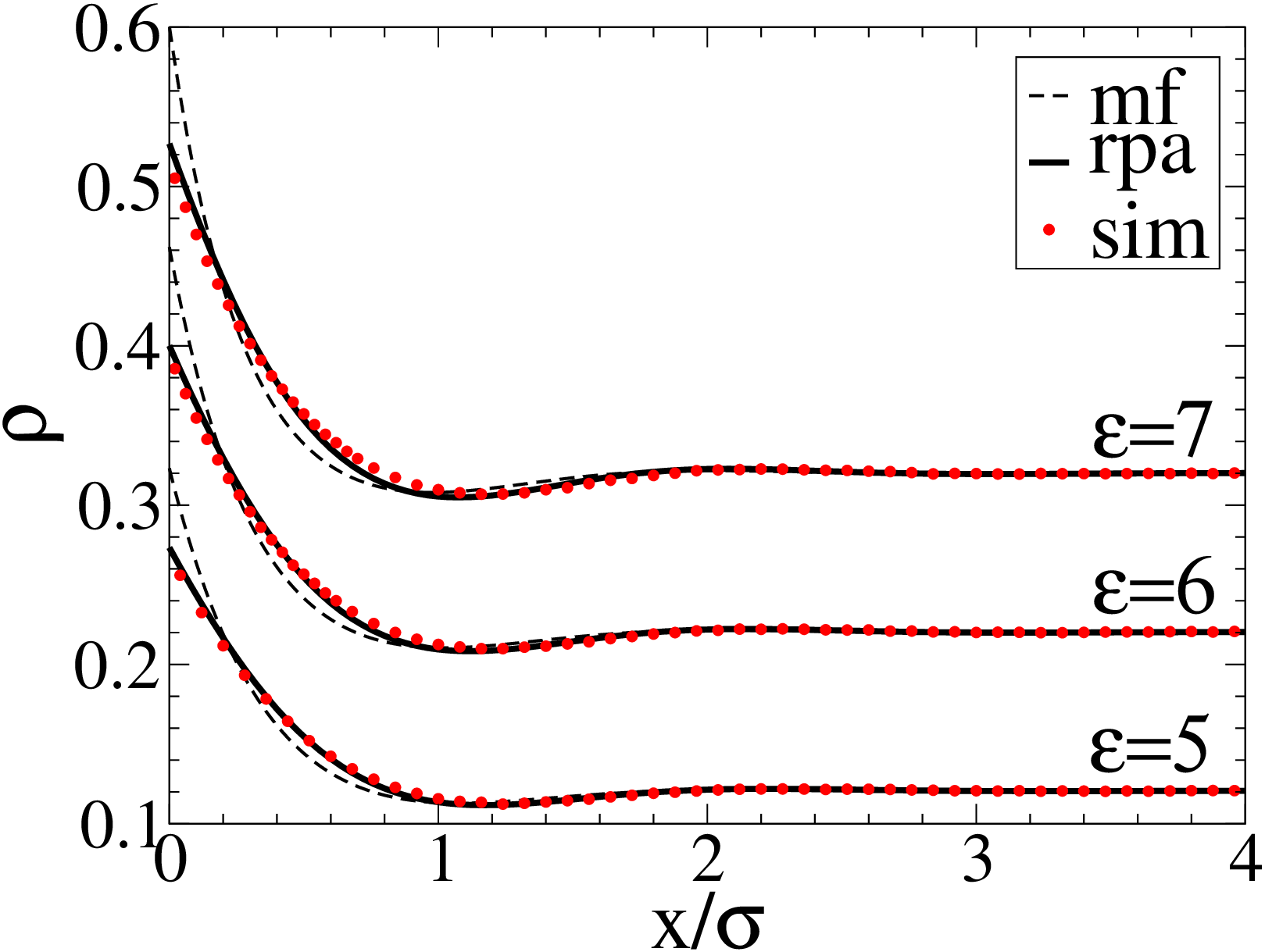}\\
 \end{tabular}
 \end{center}
\caption{Density profile for the GCM near a planar wall for different values of the interaction strength: 
$\varepsilon=5,6,7$.  A simulation box dimensions are $20\sigma\!:\!20\sigma\!:\!20\sigma$ 
and it contains $1000$ particles.  Without hard wall constraint this corresponds to a bulk reduced 
density $\eta_b\approx0.7$.  The numerical data points for the mean-field and the RPA correspond
to the same conditions.  }
\label{fig:eta_2}
\end{figure}

In Fig. (\ref{fig:eta_2}) we plot the entire density profiles near a planar wall at $x=0$.  The mean-field is not
expected to be accurate for $\varepsilon>1$ and the largest deviations from the exact results occur 
near the wall.   The RPA profile not only improves the contact region but an entire profile, even for as 
large values of the interaction strength as $\varepsilon=7$.
\graphicspath{{figures/}}

\begin{figure}[h] 
 \begin{center}
 \begin{tabular}{rr}
  \includegraphics[height=0.25\textwidth,width=0.35\textwidth]{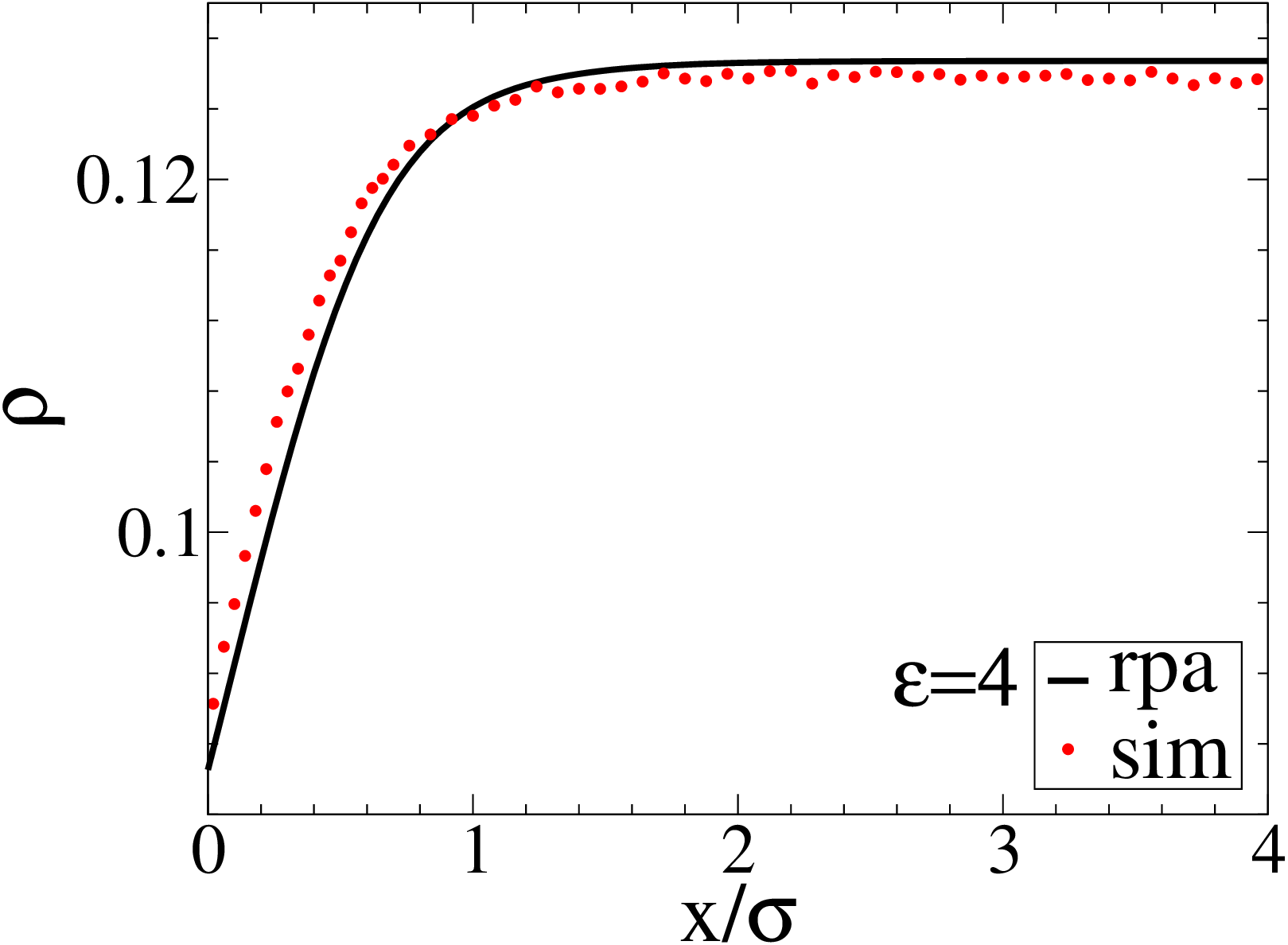}\\
 \end{tabular}
 \end{center}
\caption{The total density profile near a planar wall for the two component GCM fluid, for 
$\varepsilon=4$.  The system size and the number of particles is the same as in Fig. (\ref{fig:eta_2}), 
except the total density is made up of two species which have identical density profiles.  The density
is determined by correlations alone as the mean-field effects cancel out.  }
\label{fig:eta_3}
\end{figure}

To complete the analysis, we consider next a two component GCM system with interactions
\[ u_{ij}(x) = \left\{ 
  \begin{array}{r l}
     \varepsilon e^{-r^2/\sigma^2}, & \quad \text{if $i=j$}\\
    -\varepsilon e^{-r^2/\sigma^2}, & \quad \text{if $i\ne j$.}
\label{eq:u}
  \end{array} \right.\]
The bulk density of both species is the same, $\rho_b=\rho_b^+=\rho_b^-$, so that the mean-field contributions 
are canceled out and the density profile is determined strictly by correlations.  
In Fig. (\ref{fig:eta_3}) we compare a density profile of the RPA approximation with that from the simulation.   
The depletion of particles from the interface region is caused by unfavorable energy cost when a particle
is removed from a bulk, which requires breaking of various "bonds" with its neighbors.  
\graphicspath{{figures/}}

\section{One-component plasma}
\label{sec:OCP}
We consider next Coulomb particles and transform the previously obtained expressions of the RPA
approximation to more familiar expressions in terms of an electrostatic potential.  The resulting expressions 
are the same as those obtained for a variational Gaussian approximation within the field-theoretical 
framework \cite{Netz03,Wang10} and without resorting to the Hubbard-Stratonovich transformation.  

Coulomb charges $q$ interact via the following pair potential, 
\be
\ua = \frac{q^2}{4\pi\epsilon|{\bf r}-{\bf r}'|},
\label{eq:uc1}
\ee
where $\epsilon$ is the background dielectric constant.  
A number density of Coulomb charges, using Eq. (\ref{eq:rho_1}), is 
\be
\ra = \rho_b e^{-\beta q\psi({\bf r})}e^{\frac{1}{2}\big[h^{\rm }({\bf r},{\bf r}) - h_b(0)\big]},
\label{eq:rho_ocp}
\ee
where the external potential, in electrostatic problems generated by permanent charges 
distributed over surfaces and accounted for by the boundary conditions, is omitted from the
expression.  Furthermore, we introduce an electrostatic potential, $\psi({\bf r})$, defined as
\be
q\psi({\bf r}) = \intb\,\rb\ua.  
\label{eq:psi}
\ee
To transform the OZ-RPA equation in Eq. (\ref{eq:OZ_RPA2}) into desired form, we apply the Laplacian 
operator to both sides of the equation, 
\be
\nabla^2h^{\rm }({\bf r},{\bf r}') =  
\frac{\beta q^2}{\epsilon}\bigg[\delta({\bf r}-{\bf r}') + \ra h^{\rm }({\bf r},{\bf r}') \bigg],
\label{eq:OZ_RPA3}
\ee
where we used the identity
\be
\nabla^2\ua = -\bigg(\frac{q^2}{\epsilon}\bigg)\delta({\bf r}-{\bf r}').
\label{eq:uc2}
\ee
We carry out the same operation on Eq. (\ref{eq:psi}), 
\be
\epsilon\nabla^2\psi({\bf r}) = 
-q\rho({\bf r}),
\label{eq:poisson}
\ee
and the result is the standard Poisson equation.  

Eq. (\ref{eq:rho_ocp}), (\ref{eq:OZ_RPA3}), and (\ref{eq:poisson}) constitute the RPA approximation
for a density distribution and an electrostatic potential within the RPA level of approximation. 
Correlational contributions enter through the correlations in the number density $h^{\rm }({\bf r},{\bf r}')$.  
These can be related to correlations in electrostatic potential using a slightly rearranged OZ-RPA equation, 
\ba
&&h^{\rm }({\bf r},{\bf r}') = \nonumber\\
&&-\beta\int d{\bf r}''\,\bigg[\rc h^{\rm }({\bf r}',{\bf r}'')+\delta({\bf r}'-{\bf r}'')\bigg]u({\bf r},{\bf r}'').\nonumber\\
\label{eq:OZ_hole}
\ea
The term in square brackets, $\rho({\bf r}'')h^{\rm }({\bf r}'',{\bf r}')$, is identified as a correlation hole 
generated by a fixed particle at ${\bf r}'$ and the delta function denotes the density of a fixed particle.  
The integral on the right hand side can be reinterpretted as a perturbation
of an electrostatic potential, $\Psi_{}({\bf r},{\bf r}')$, caused by a fixed particle at ${\bf r}'$ (the 
total electrostatic potential is $\psi({\bf r})+\Psi_{}({\bf r},{\bf r}')$).  The OZ-RPA equation simply becomes
\be
h_{}({\bf r},{\bf r}') = -\beta q\Psi_{}({\bf r},{\bf r}'),
\label{eq:h_psi}
\ee
and the proportionality between the two fluctuating quantities is established.  Note that this is not an exact
equality but a result specific of the RPA approximation.  


For a concrete example we consider a counterion only system confined to a half-space $x>0$.  
The counterion charge is $q=e$.
The wall surface charge $\sigma_c$ at $x=0$ assures neutrality of the system.  The dielectric 
constant is the same on both sides of the wall.  As the bulk density far away from the wall vanishes, 
the contact density is determined solely by the surface charge (not the pressure), 
\be
\rho(0) = -\int_0^{\infty}dz\,\rho(z)\frac{\partial U(z)}{\partial z} = \frac{\beta\sigma_c^2}{2\epsilon},
\ee
where $-\partial U(z)/\partial z=-e\sigma_c/2$ is a constant force felt by particles on account of a uniform 
wall charge.  The mean-field solution to this problem is
\be
\rho_{\rm mf}(x) 
= \frac{\beta\sigma_c^2}{2\epsilon}\bigg[\frac{1}{1+\beta q\sigma_c x/2\epsilon}\bigg]^2,
\ee
and it captures a weakly correlated limit.  On the opposite end is the strong-coupling limit
\cite{Netz00b,Trizac11}, 
\be
\rho_{\rm sc}(x) = \frac{\beta\sigma_c^2}{2\epsilon}e^{-\beta q\sigma_c x/2\epsilon}.
\ee

\begin{figure}[h] 
 \begin{center}
 \begin{tabular}{rr}
  \includegraphics[height=0.2\textwidth,width=0.35\textwidth]{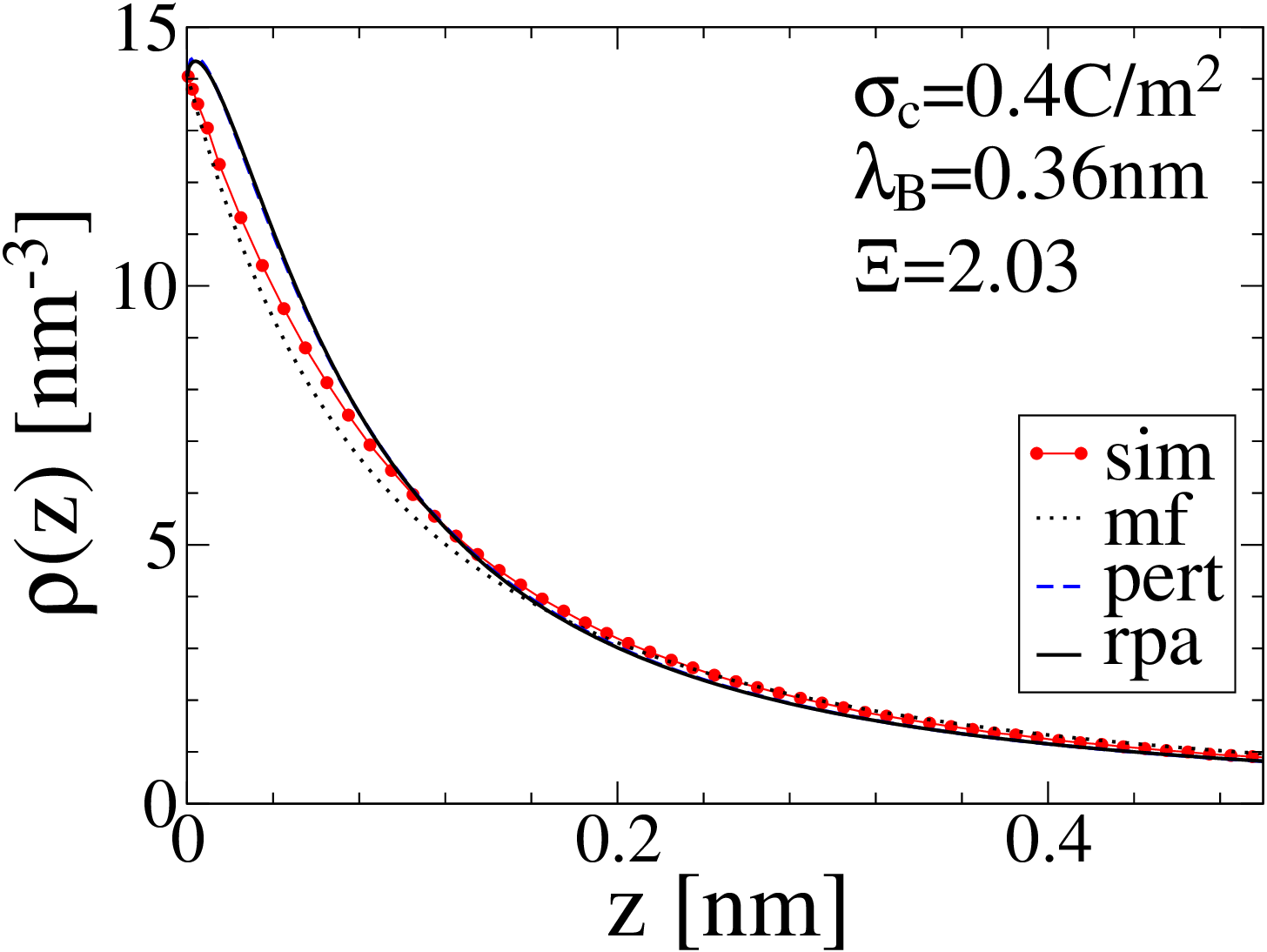}\\
  \includegraphics[height=0.2\textwidth,width=0.35\textwidth]{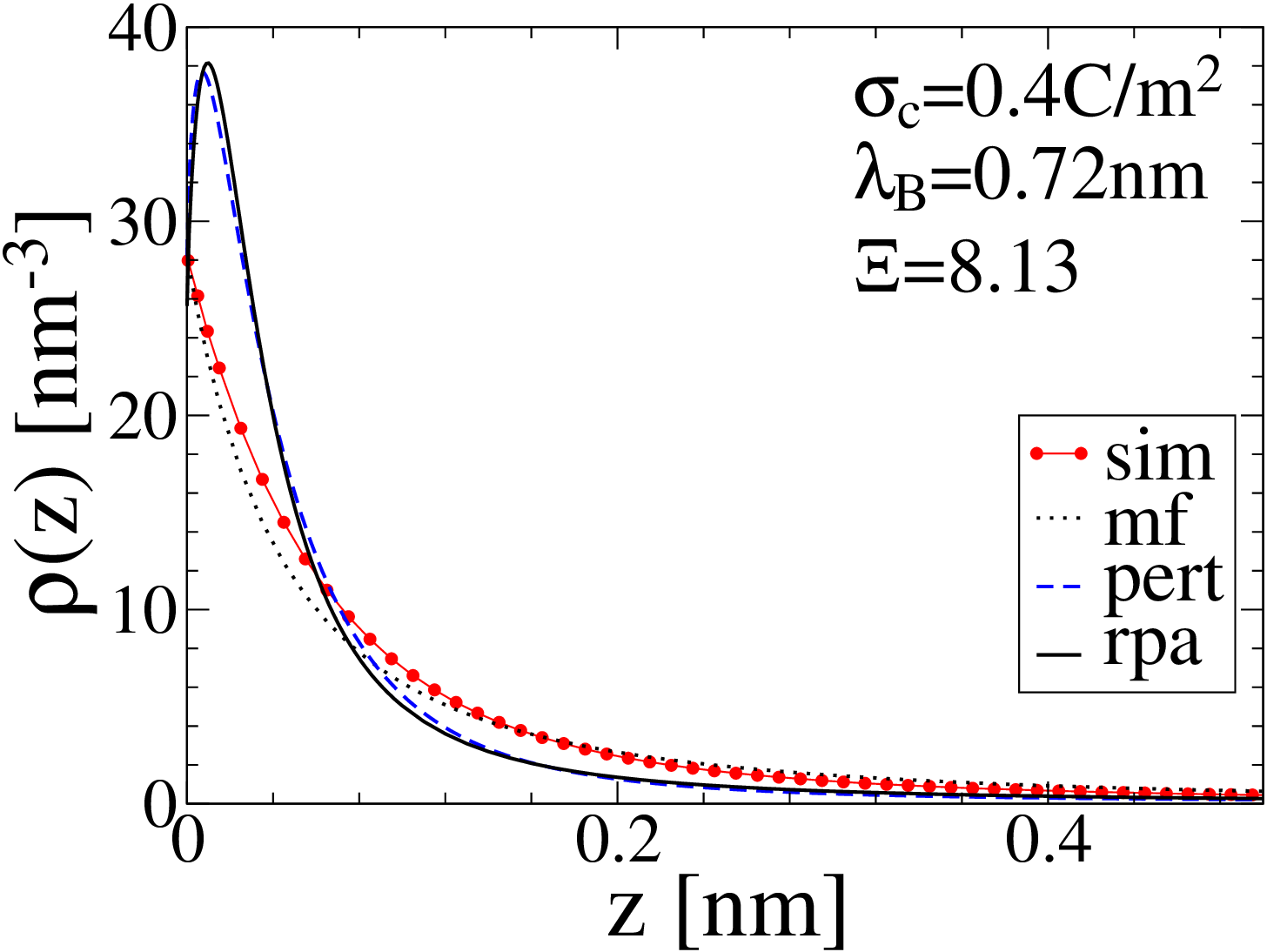}\\
 \end{tabular}
 \end{center}
\caption{Counterion density profiles for a counterion-only wall model.  $\Xi$ denotes the 
coupling constant corresponding to the ratio of the Bjerrum and the Gouy-Chapman length, $\lambda_B$ 
and $1/(2\pi\lambda_B\sigma_c)$, respectively.  The larger the $\Xi$, the more significant the correlations. }
\label{fig:rho_ocp}
\end{figure}

As correlations become significant, the density evolves from one functional form to another, 
$\rho_{\rm mf}\to\rho_{\rm sc}$.  A perturbative Gaussian approach for a counterion-only system
yields a semi-analytic expression for a density correction $\Delta\rho(z)$,  
$\rho(z)=\rho_{\rm mf}(z)+\Delta\rho(z)$ \cite{Netz00,Samaj13}, where
$
\int_0^{\infty}dz\,\Delta\rho(z) = 0,
$
to maintain neutrality, and $\Delta\rho(0)=0$, not to violate the contact value theorem.  The corrected 
density develops a "hump" at a short distance from a wall.  It is difficult to justify or trace the hump to 
a physical cause as it is not confirmed by simulations;  simulations always yield a non-monotonic density 
profile.  It should be concluded that the evolution of the "hump" is an artifact of the 
approximation.  It is hoped that the self-consistent equations of the RPA method eliminate the 
"hump" in iterative steps and in closer agreement with the true system.  Our computations, however,
indicate that self-consistency only slightly alters the results of the perturbative approach (see Fig. \ref{fig:rho_ocp}). 
\graphicspath{{figures/}}
The conclusion is that the RPA approximation is not very accurate for the counterion-only system, 
and modifies the perturbative results only negligibly.


\section{Electrolyte}
\label{sec:electrolyte}
Continuing with electrostatics, we move toward electrolytes.  
We consider a symmetric electrolyte, $q:q$, with a bulk concentration of both species $\rho_b$.  
The two types of pair interactions are
\[ u_{kl}({\bf r},{\bf r}') = \left\{ 
  \begin{array}{l l}
    \,\,~u({\bf r},{\bf r}') & \quad \text{if $k=l$}\\
    -u({\bf r},{\bf r}') & \quad \text{if $k\ne l$},
  \end{array} \right.\]
where $\ua$ is the Coulomb potential given in Eq. (\ref{eq:uc1}).

For a two species system, there are four different correlation functions, $h_{kl}({\bf r},{\bf r}')$.  
Within the RPA they can be expressed in terms of a single function.  
Accordingly, we have
\[ h_{kl}^{\rm }({\bf r},{\bf r}') = \left\{ 
  \begin{array}{l l}
    \,\,~h^{\rm }({\bf r},{\bf r}') & \quad \text{if $k=l$}\\
    -h^{\rm }({\bf r},{\bf r}') & \quad \text{if $k\ne l$}
  \end{array} \right.\]
and $h^{\rm }({\bf r},{\bf r}')$ is obtained from the OZ-RPA relation,
\ba
&&h^{\rm }({\bf r },{\bf r}') = \nonumber\\
&&-\beta\intc\,
\bigg[\rho({\bf r}'')h^{\rm }({\bf r}',{\bf r}'') + \delta({\bf r}'-{\bf r}'')\bigg]u({\bf r},{\bf r}''),\nonumber\\
\label{eq:OZ_RPA4}
\ea
where $\ra = \rho_+({\bf r}) + \rho_-({\bf r})$
is the total density.  Within the RPA, the number density of each species is
\be
\rho_{\pm}({\bf r}) = 
\rho_b e^{\mp\beta q\psi({\bf r})}e^{\frac{1}{2}\big[h^{\rm }({\bf r},{\bf r})-h_b^{\rm }(0)\big]},
\ee
where the correlation function is obtained from a transformed Eq. (\ref{eq:OZ_RPA4}) (by applying 
the Laplacian operator to both sides of the OZ-RPA equation), 
\be
\epsilon\nabla^2h^{\rm }({\bf r},{\bf r}') 
= \beta q^2 \bigg[\rho({\bf r})h^{\rm }({\bf r},{\bf r}') + \delta({\bf r}-{\bf r}')\bigg].  
\label{eq:h_el}
\ee
Together with the Poisson equation,
\be
\epsilon\nabla^2\psi({\bf r}) = -\rho_c({\bf r}),
\ee
where $\rho_c({\bf r}) = q\rho_+({\bf r}) - q\rho_-({\bf r})$ is a charge density, 
we have a complete approximation for a density and electrostatic potential of a symmetric electrolyte.   

As in the case of a one-component plasma, we may link the correlations in a number density to the 
correlations in an electrostatic potential.  We identify the term in brackets in Eq. (\ref{eq:OZ_RPA4}) as a 
charge correlation hole generated by fixing either a positive or a negative charge $q$ at ${\bf r}'$.  
For a fixed positive charge we have
\ba
\rho_{\rm hole}({\bf r},{\bf r}') &=& \rho_+({\bf r})h_{++}({\bf r},{\bf r}') - \rho_{-}({\bf r})h_{+-}({\bf r},{\bf r}')
\nonumber\\&=&\rho_+({\bf r})h_{}({\bf r},{\bf r}') + \rho_{-}({\bf r})h_{}({\bf r},{\bf r}')
\nonumber\\&=&\ra\ha. 
\ea
Consequently, the fluctuations in the number density are proportional to the fluctuations in electrostatic
potential, 
\be
h^{\rm }({\bf r},{\bf r}') = -\beta q\Psi({\bf r},{\bf r}'),
\ee
as was previously demonstrated for a one-component plasma in Eq. (\ref{eq:h_psi}).

Note that the excess chemical potential does not include the mean-field contributions and depends
exclusively on correlations, 
\be
\mu_{\rm ex}^{\pm} = -\frac{1}{2}\lim_{r\to 0}\bigg[k_BTh_b^{\rm }(r)  + u(r)\bigg].
\ee
The mean-field contributions cancel out by virtue of charge neutrality.  
The same is true of pressure which reads 
\ba
P_{\rm ex} &=& 
- \frac{k_BT\rho_b}{2}\lim_{r\to 0} \bigg[h_{b}^{\rm }(r)
-\int_0^{1} d{\lambda}\,\frac{h_b^{\lambda}(r)}{\lambda}\bigg],\nonumber\\
\ea
and where $\rho_b=\rho^b_{+}+\rho_-^b$ is the bulk total density.  

For bulk electrolytes Eq. (\ref{eq:h_el}) recovers the Debye-H\"uckel theory for a point-charge,  
\be
\frac{d^2h_{\lambda}^b(r)}{dr^2} = \kappa_{\lambda}^2h^b_{\lambda}(r) + \bigg(\frac{\lambda\beta q^2}{\epsilon}\bigg)\delta(r).  
\ee
where $\kappa_{\lambda}=\sqrt{\lambda\beta q^2\rho_b/\epsilon}$ is the screening parameter.  The Debye-H\"uckel solution is
\be
h_b^{\lambda}(r) = -\frac{\lambda \beta q^2 e^{-\kappa_{\lambda} r}}{4\pi\epsilon r}, 
\ee
and the excess pressure becomes
\ba
P_{\rm ex} &=& \frac{q^2\rho_b}{8\pi\epsilon}\lim_{r\to 0} \bigg[\frac{e^{-\kappa r}}{r}
-\int_0^{1} d{\lambda}\,\frac{e^{-\kappa_{\lambda} r}}{r}\bigg]\nonumber\\
&=& -\frac{\kappa}{3}\frac{q^2\rho_b}{8\pi\epsilon} 
\ea
where $\kappa\equiv\kappa_{\lambda=1}$.  
The total pressure may be written as \cite{Yan02}
\be
\beta P = \rho_b - \frac{\kappa^3}{24\pi}.  
\ee
We note that the RPA correlations reduce the ideal gas pressure.  From the contact value theorem 
we may infer that at neutral interfaces 
the contact density will be lower than that in a bulk, indicating a depletion zone that is not caused by 
dielectric discontinuity but a more efficient bonding arrangement within a bulk.

\begin{figure}[h] 
 \begin{center}
 \begin{tabular}{rr}
  \includegraphics[height=0.2\textwidth,width=0.35\textwidth]{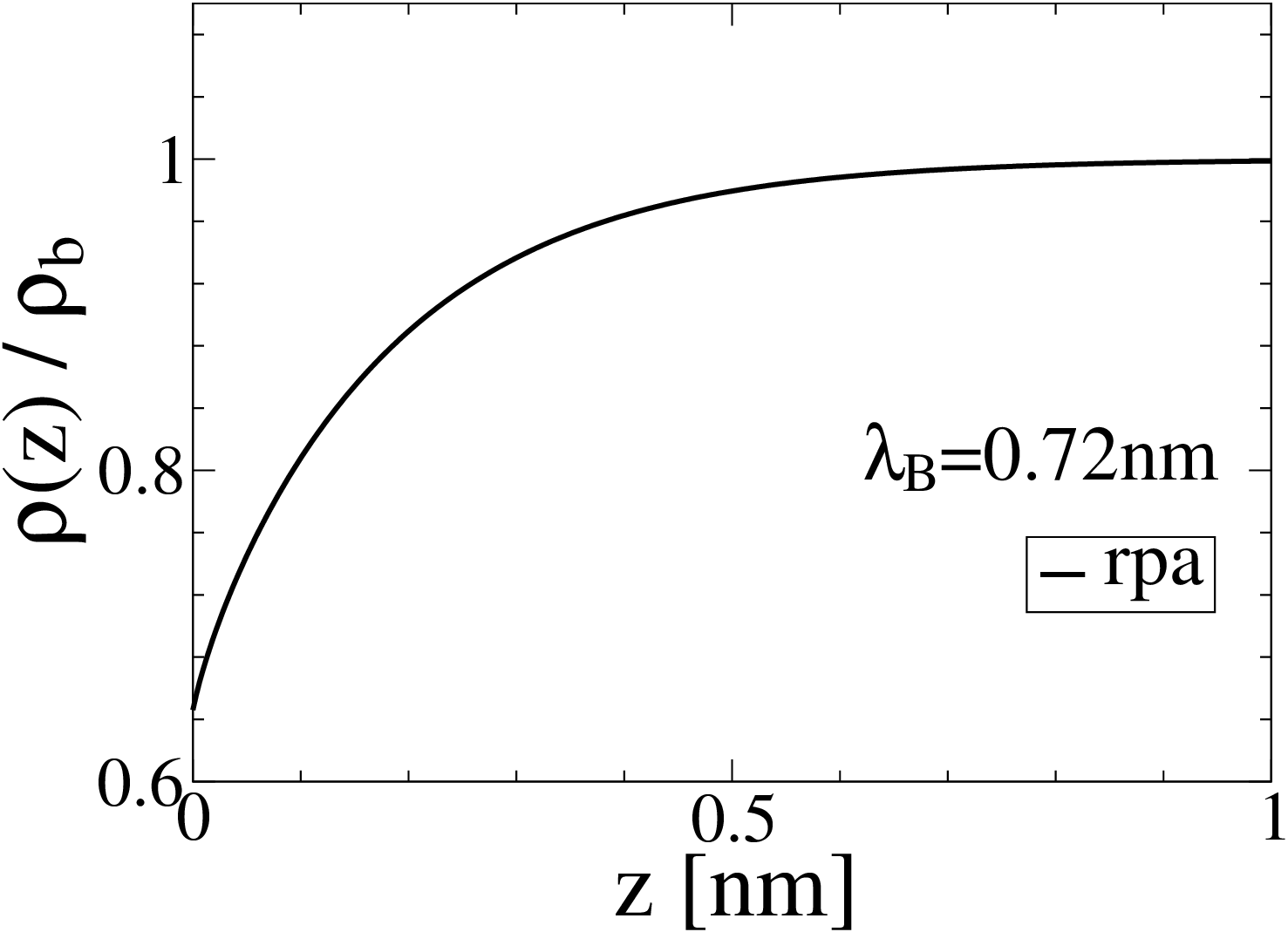}\\
 \end{tabular}
 \end{center}
\caption{Density profiles for a 1:1 electrolyte near a wall at $z=0$.  The Bjerrum 
and the screening lengths are
$\lambda_B=0.72\,{\rm nm}$ 
and $\kappa^{-1}=0.33\,{\rm nm}$, respectively.  The ions are depleted from a wall region to minimize the number of particles at 
an interface.  
}
\label{fig:rho_11}
\end{figure}

Once again resort to a simple wall geometry for testing purposes.  The wall is uncharged and its only function 
is to create an interface between an electrolyte and an empty space.  Dielectric constant is uniform
across the interface and everywhere else.  
The results in Fig. (\ref{fig:rho_11}) indicate the depletion zone near an interface, neither caused by dielectric 
discontinuity nor finite ion size but correlations.  The repulsion of ions from an interface in turn increases
the surface tension \cite{Yan01}. 
As for the two component GCM in Fig. (\ref{fig:eta_3}) this is caused by a more efficient salvation of 
ions in a bulk, where each ion is accompanied by an opposite charge-cloud.  Near the interface,
the charge-cloud is deformed by a nearby interface rising the cost of an energy.   
\graphicspath{{figures/}}

\section{Conclusion}
The free energy formulation using the adiabatic connection offers a simple way for incorporating the OZ 
equation into a theoretical framework for liquids 
and may be considered as an extension of the integral equation theories to inhomogenous fluids.  
Finally, the choice of closure determines an approximation.  
In the present work we explore the RPA closure, $c_{\lambda}({\bf r},{\bf r}')=-\lambda\beta\ua$.  
We demonstrate that the resulting general RPA approximation is equivalent to the field-theoretical 
variational Gaussian approximation, derived completely within the liquid-state theory.

We test the developed RPA framework for different inhomogeneous fluids for a simple wall geometry.  
For the Gaussian core model the RPA density profiles show decisive improvement over the mean-field and
are in good agreement with a simulation, even for large interaction strengths.  For the counterion-only 
system the RPA is less accurate.  Like the perturbative Gaussian approximation \cite{Netz00,Samaj13} 
the density profile develops an unphysical bump near the wall.  In comparison, the simulated profiles 
are always monotonically decreasing.  Furthermore, the self-consistency of our approach appears to modify 
rather negligibly the profile obtained from the perturbative scheme \cite{Netz00,Samaj13}.  Consequently, 
we conclude that the RPA (or the variational Gaussian approximation) is not an accurate theoretical tool for 
strongly correlated Coulomb fluids.  Finally, we test apply the RPA to a symmetric electrolyte near
a neutral interface, without dielectric discontinuity.  We observe the depletion of density near an interface
generated exclusively by correlations, since the mean-field contributions in this system are cancelled out.

As final remarks, we restate that we did not see that the self-consistent scheme of the RPA (or the variational
Gaussian) produces significant modifications
in comparison with the results obtained perturbatively.  If there are situations where
self-consistency is crucial, we cannot be sure, but for systems and parameters considered in this work we did 
not come across such conditions.  Finally, as the future project, we think it worthwhile to explore the
adiabatic connection framework presented in this work but for more accurate closures.  Self-consistency
in these more advanced closures may turn out to be more significant.

\begin{acknowledgments}
Most computations were carried out on machines belonging to the Laboratoire de Physico-Chime Th\'eorique, ESPCI
by friendly permission of Tony Maggs.  
This research was partly supported by the Chinese National Science Foundation, the grant number 11574198.
\end{acknowledgments}




\begin{thebibliography}{99}
\bibitem{Evans79}
R. Evans, Adv. Phys. A {\bf 28}, 143 (1979). 
\bibitem{Rosenfeld89}
Y. Rosenfeld, Phys. Rev. Lett. {\bf 63}, 980 (1989).
\bibitem{Tarazona08}
P. Tarazona, J. A. Cuesta, and Y. Mart\'inez-Rat\'on, Lect. Notes Phys. {\bf 753} 247 (2008).
\bibitem{Evans09}
R. Evans, Lecture Notes at 3rd Warsaw School of Statistical Physics 
(Warsaw University Press, Kazimierz Dolny, 2009) pp. 43?85.  
\bibitem{Roth10}
R. Roth, J. Phys.: Condens. Matter {\bf 22}, 063102 (2010).
\bibitem{Frydel12}
D. Frydel, Y. Levin, J. Chem. Phys. {\bf 137}, 164703 (2012).
\bibitem{Barker67}
J. A. Barker, D. Henderson, J. Chem. Phys. {\bf 47}, 2856 (1967).
\bibitem{Schmidt99a}
M. Schmidt, J. Phys.: Condens. Matter {\bf 11}, 10163 (1999). 
\bibitem{Schmidt99b}
M. Schmidt, Phys. Rev. E {\bf 60}, R6291 (1999). 
\bibitem{Schmidt00}
M. Schmidt, Phys. Rev. E {\bf 62} 4976, (2000).
\bibitem{Lowen02}
H. L\"owen, J. Phys.: Condens. Matter {\bf 14} 11897 (2002).
\bibitem{Sweatman02}
M B Sweatman, J. Phys.: Condens. Matter {\bf 14}, 11921 (2002).


\bibitem{Hansen00}
A. A. Louis, P. G. Bolhuis, and J. P. Hansen, Phys. Rev. E {\bf 62}, 7961 (2000).
\bibitem{David07}
A. Abrashkin, D. Andelman, and H. Orland, Phys. Rev. Lett. {\bf 99}, 077801 (2007).
\bibitem{Frydel11}
D. Frydel, J. Chem. Phys. {\bf 134}, 234704 (2011).
\bibitem{Frydel13}
D. Frydel and Y. Levin, J. Chem. Phys. {\bf 138}, 174901 (2013).
\bibitem{Frydel15a}
http://arxiv.org/abs/1411.7577
\bibitem{Frydel15}
D. Frydel, Eur. J. Phys. {\bf 36}, 065050 (2015).

\bibitem{Rudi88}
R. Podgornik, B. Zeks,  J. Chem. Soc., Faraday Trans. 2, {\bf 84}, 611 (1988).
\bibitem{Attard88}
P. Attard, D. J. Mitchell, B. W. Ninham, J. Chem. Phys. {\bf 88} 4987 (1988).
\bibitem{Duncan92}
R. D. Coalson, A. Duncan, J. Chern. Phys., {\bf 97}, 205653 (1992).
\bibitem{Netz00}
R. R. Netz and H. Orland, Europhys. J. E {\bf 1}, 67 (2000).
\bibitem{Netz03}
R. Netz, H. Orland, Eur. Phys. J. E {\bf 11}, 301 (2003).
\bibitem{Wang10}
Z.-G Wang, Phys. Rev. E {\bf 81}, 021501 (2010).
\bibitem{Hatlo14}
P. Duncan, M, M, Hatlo, L. Lue, 
{\it A Field-Theory approach for modeling electrostatic interactions in soft matter},
in “Proceedings of the CECAM Workshop” 
New Challenges in Electrostatics of Soft and Disordered Matter (Pan Stanford, 2014). 
\bibitem{Tony14}
Zhenli Xu, A. C. Maggs,  J. Comp. Phys. {\bf 275},  (2014).

\bibitem{Pines52}
D. Pines and D. Bohm, Phys. Rev. {\bf 85}, 338 (1952).
\bibitem{Mann57}
M. Gell-Mann, K.A. Brueckner, Phys. Rev. 106, 364 (1957).
\bibitem{Perdew77}
J. P. Perdew and D. C. Langreth, Phys. Rev. B {\bf 15}, 2884 (1977).
\bibitem{Ren12}
X. Ren, P. Rinke, C. Joas, M. Scheffler, J. Mater. Sci. {\bf 47}, 7447 (2012).

\bibitem{Netz00b}
A. G. Moreira and R. R. Netz, Europhys. Lett. {bf 52}, 705 (2000).
\bibitem{Trizac11}
L. \^Samaj and E. Trizac, {\sl Phys. Rev. Lett.}, {\bf 106}, 078301 (2011).














\bibitem{Samaj13}
L. \^Samaj, Eur. Phys. J. E {\bf 36}, 100 (2013).  



\bibitem{Yan02}
Y. Levin, Rep. Prog. Phys. {\bf 65}, 1577 (2002).

\bibitem{Yan01}
Y. Levin and J. E. Flores-Mena, Eur. Phys. J. {\bf 56},  187 (2001).



\end{thebibliography}
\end{document}